# FINDCHIRP: An algorithm for detection of gravitational waves from inspiraling compact binaries

Bruce Allen,[1,2] Warren G. Anderson,[1] Patrick R. Brady,[1] Duncan A. Brown,[1,3,4,5] and Jolien D. E. Creighton[1]

[1]*Department of Physics, University of Wisconsin–Milwaukee, P.O. Box 413, Milwaukee, Wisconsin 53201, USA*
[2]*Max-Planck-Institut für Gravitationsphysik (Albert-Einstein-Institut), Callinstraße 38, D-30167 Hannover, Germany*
[3]*Theoretical Astrophysics 130-33, California Institute of Technology, Pasadena, California 91125, USA*
[4]*LIGO Laboratory, California Institute of Technology, Pasadena, California 91125, USA*
[5]*Department of Physics, Syracuse University, Syracuse, New York 13244, USA*

Matched-filter searches for gravitational waves from coalescing compact binaries by the LIGO Scientific Collaboration use the FINDCHIRP algorithm: an implementation of the optimal filter with innovations to account for unknown signal parameters and to improve performance on detector data that has nonstationary and non-Gaussian artifacts. We provide details on the FINDCHIRP algorithm as used in the search for subsolar mass binaries, binary neutron stars, neutron star–black hole binaries, and binary black holes.

## I. INTRODUCTION

For the detection of a known signal it is well-known that, in the presence of stationary and Gaussian noise, the use of a matched filter is the optimal detection strategy [1]. Such signals are the anticipated gravitational waveforms from the inspiral of binary systems containing compact objects, i.e., binaries whose components are neutron stars and/or black holes [2,3]. These gravitational-wave sources are among the most promising targets for ground-based gravitational-wave detectors, such as the Laser Interferometer Gravitational-wave Observatory (LIGO) [4] and Virgo [5].

Some practical complications arise for the gravitational-wave detection problem, however, because: (i) the signal is not precisely known—it is parametrized by the binary companions' masses, an initial phase, the time of arrival, and various parameters describing the distance and orientation of the system relative to the detector that can be combined into a single parameter we call the "effective distance," and (ii) the detector noise is not perfectly described as a stationary Gaussian process. Standard techniques for extending the simple matched filter to search over the unknown parameters involve using a quadrature sum of matched-filter outputs for orthogonal-phase waveforms (thereby eliminating the unknown phase), use of Fourier transform to efficiently apply the matched filters for different times of arrival, and use of a bank of templates to cover the parameter space of binary companion masses [6–12]. Methods for making the matched filter more robust against non-Gaussian noise artifacts, e.g., by examining the relative contributions of frequency-band-limited matched-filter outputs (vetoing those transients that produce large matched-filter outputs but have a time-frequency decomposition that is inconsistent with the expected waveform), have also been explored [13] and found to be effective.

The FINDCHIRP algorithm is the implementation of matched filtering used by the LIGO Scientific Collaboration (LSC) and Virgo's offline searches for gravitational waves from low-mass ($2M_\odot < M = m_1 + m_2 < 35M_\odot$), high-mass ($25M_\odot < M < 100M_\odot$), and primordial black hole ($0.2M_\odot < M < 2M_\odot$) coalescing binaries [14–25]. FINDCHIRP has also been used to search for supermassive black holes in data from the LISA Mock Data Challenge [26–28] and for comparisons with numerical relativity waveforms for both high-mass and low-mass binaries [29]. Several aspects of the algorithm have been described in passing in the above references, but here we provide a detailed and comprehensive description of our algorithm as used in the LSC and Virgo's search for coalescing compact binaries.

The FINDCHIRP algorithm is the part of the search that (i) computes the matched-filter response to the interferometer data for each template in a bank of templates, (ii) computes a chi-squared discriminant [13] (if needed) to reject instrumental artifacts that produce large spurious responses of the matched filter but otherwise do not resemble an expected signal, and (iii) selects candidate events or triggers based on the matched-filter and chi-squared outputs. This is a fundamental part of the search for coalescing compact binaries, but the search also consists of several other important steps such as data selection and conditioning, template bank generation, rejection of candidate events by vetoes based on auxiliary instrumental channels and multidetector coincidence, and computation of the false alarm rate of candidate triggers.

The entire search pipeline, which is a transformation of raw interferometer data into candidate events, contains all these aspects. The specific details of the pipeline used to search for gravitational waves depends on the target population (e.g., a triggered search for gamma-ray bursts or an all-sky search for low-mass compact binaries) and the particular "science run" of the LIGO and/or Virgo detectors. We refer to Refs. [14–16,18,19,21–25,30–32] for descriptions of the pipelines that have used the algorithms described here. This paper is not intended to provide documentation for our implementation of the FINDCHIRP algorithm. (This can be found in Refs. [33,34].) Rather, this paper is intended to describe the algorithm itself.

## II. NOTATION

Our conventions for the Fourier transform are as follows. For continuous quantities, the forward and inverse Fourier transforms are given by

$$\tilde{x}(f) = \int_{-\infty}^{\infty} x(t) e^{-2\pi i f t} dt, \quad (2.1a)$$

and

$$x(t) = \int_{-\infty}^{\infty} \tilde{x}(f) e^{2\pi i f t} df, \quad (2.1b)$$

respectively, so $\tilde{x}(f)$ is the Fourier transform of $x(t)$. If these continuous quantities are discretized so that $x[j] = x(j\Delta t)$ where $1/\Delta t$ is the sampling rate and $j = 0, \ldots, N-1$ are $N$ sample points, then the discrete approximation to the forward and inverse Fourier transforms are

$$\tilde{x}[k] = \Delta t \sum_{j=0}^{N-1} x[j] e^{-2\pi i j k/N}, \quad (2.2a)$$

$$x[j] = \Delta f \sum_{k=0}^{N-1} \tilde{x}[k] e^{2\pi i j k/N}, \quad (2.2b)$$

where $\Delta f = 1/(N\Delta t)$ and $\tilde{x}[k]$ is an approximation to the value of the continuous Fourier transform at frequency $k\Delta f$: $\tilde{x}[k] \approx \tilde{x}(k\Delta f)$ for $0 \leq k \leq \lfloor N/2 \rfloor$ and $\tilde{x}[k] \approx \tilde{x}((k-N)\Delta f)$ for $\lfloor N/2 \rfloor < k < N$ (negative frequencies). Here $\lfloor a \rfloor$ means the greatest integer less than or equal to $a$. The DC component is $k = 0$ and, when $N$ is even, $k = N/2$ corresponds to the Nyquist-frequency.

Notice that our convention is to have the Fourier components $\tilde{x}[k]$ normalized so as to have the same units as the continuous Fourier transform $\tilde{x}(f)$, i.e., the discretized versions of the continuous forward and inverse Fourier transforms carry the normalization constants $\Delta t$ and $\Delta f$ respectively. Numerical packages instead compute the discrete Fourier transform (DFT)

$$y[k] = \sum_{j=0}^{N-1} x[j] e^{\mp 2\pi i j k/N} \quad (2.3)$$

where the minus sign in the exponential refers to the forward DFT and the positive sign refers to the *reverse*[1] DFT. The DFT is efficiently implemented via the fast

---

[1]We use the term *reverse* rather than *inverse* since the inverse would include an overall normalization factor of $1/N$.

Fourier transform (FFT) algorithm. Thus, it is important to write the most computationally demanding terms in the form of Eq. (2.3) so that this computation can be done most efficiently.

Throughout this paper we will reserve the indices $j$ to be a time index (which labels a particular time sample), $k$ to be a frequency index (which labels a particular frequency bin), $m$ to be an index over a bank of templates, and $n$ to be an index over analysis segments. Thus, for example, the quantity $z_{m,n}[j]$ will be the $j$th sample of analysis segment $n$ of the matched-filter output for the $m$th template, and $\tilde{z}_{m,n}[k]$ will be the $k$th frequency bin of the Fourier transform of the matched-filter output for the same template and analysis segment.

## III. WAVEFORM

For first-generation gravitational-wave detectors (such as Initial LIGO), the gravitational-wave signal from a compact binary inspiral waveform can be accurately modeled by the restricted post-Newtonian waveform [35] below approximately $M \sim 12 M_\odot$ [36] for nonspinning bodies. At higher masses, resummation techniques such as the effective one body (EOB) waveforms better reproduce the waveforms computed by numerically solving the full nonlinear Einstein equations [37,38]. In either case, the two polarizations of the gravitational-wave produced by such a system exhibit a monotonically-increasing frequency and amplitude as the orbital motion radiates away energy and decays. The waveform, often called a *chirp* waveform, is given for $t < t_c$ by

$$h_+(t) = -\frac{1 + \cos^2\iota}{2} \left(\frac{G\mathcal{M}}{c^2 D}\right) \left(\frac{t_c - t}{5G\mathcal{M}/c^3}\right)^{-1/4}$$
$$\times \cos[2\phi_c + 2\phi(t - t_c; M, \mu)], \quad (3.1a)$$

$$h_\times(t) = -\cos\iota \left(\frac{G\mathcal{M}}{c^2 D}\right) \left(\frac{t_c - t}{5G\mathcal{M}/c^3}\right)^{-1/4}$$
$$\times \sin[2\phi_c + 2\phi(t - t_c; M, \mu)], \quad (3.1b)$$

where $D$ is the distance from the source, $\iota$ is the angle between the direction to the observer and the orbital angular momentum axis of the binary system, $\mathcal{M} = \mu^{3/5} M^{2/5} = \eta^{3/5} M$ (where $M = m_1 + m_2$ is the total mass of the two companions, the reduced mass is $\mu = m_1 m_2 / M$, and $\eta = \mu/M$) is the *chirp mass*, and $\phi(t - t_c; M, \mu)$ is the orbital phase of the binary (whose evolution also depends on the individual masses of the binary companions) [39,40]. Here, $t_c$ and $\phi_c$ are the time and phase of the binary coalescence when the waveform is terminated, known as the *coalescence time* and *coalescence phase* $\phi(0; M, \mu) = 0$. Note that in Ref. [39] the integration constant $t_c$ and $\phi_c$ are implicitly contained in the orbital phase $\phi(t)$. We have chosen to express them explicitly.

The coalescence phase and time may lack a formal definition (as in the case of numerical waveforms),

however for a *detection* algorithm, the formal definition of the coalescence time is not critical, as long as any offset caused by the choice of coalescence time is constant between the detectors in the network. For restricted post-Newtonian waveforms, we define $t_c$ to be the time at which the gravitational-wave frequency becomes infinite within the post-Newtonian formalism. The Initial LIGO science runs (S1–S5) used the second-post-Newtonian waveform, whereas the Enhanced LIGO (S6) analysis uses post-Newtonian waveforms computed to 3.5th order. In this paper we illustrate the FINDCHIRP algorithm using second-order post-Newtonian waveforms. The extension of the algorithm to EOB waveforms (and other waveforms computed in the time-domain) is described in Appendix A and the extension to higher post-Newtonian orders is described in Appendix E. We assume that the binary system's center-of-mass is at rest with respect to the detector frame (otherwise this motion "redshifts" the binary's masses). We also assume that the time that the signal is in the detector's sensitive bandwidth is short compared to 24h, so that the detector does not change orientation significantly as the earth rotates.

The gravitational-wave strain induced in a particular detector depends on the detector's antenna response to the two polarizations of the gravitational waveform. The induced strain on the detector is given by

$$h(t) = F_+(\alpha, \delta, \psi, t) h_+(t + t_c - t_0) \\ + F_\times(\alpha, \delta, \psi, t) h_\times(t + t_c - t_0) \quad (3.2)$$

where $t_0$ is the *termination time* (the time at the detector at which the coalescence occurs, i.e., the detector time when the gravitational-wave frequency, according to restricted post-Newtonian approximation, becomes infinite), $t_0 - t_c$ is the propogation time from the source to the detector, and $F_+$ and $F_\times$ are the antenna response functions for the incident signal; these functions depend on the location of the source with respect to the reference frame of the detector, which is described by the right ascension and declination of the source, $(\alpha, \delta)$, the arrival time at the detector, $t_0$, and on the polarization angle $\psi$ [41]. The antenna response functions are very nearly constant in time over the duration of the short inspiral signal. Thus the strain on a particular detector can be written as

$$h(t) = -\left(\frac{G\mathcal{M}}{c^2 D_{\text{eff}}}\right)\left(\frac{t_0 - t}{5G\mathcal{M}/c^3}\right)^{-1/4} \\ \times \cos[2\phi_0 + 2\phi(t - t_0; M, \mu)], \quad (3.3a)$$

where $\phi_0$ is the *termination phase* which is related to the coalescence phase by

$$2\phi_0 = 2\phi_c - \arctan\left(\frac{F_\times}{F_+} \frac{2\cos\iota}{1 + \cos^2\iota}\right) \quad (3.3b)$$

and

$$D_{\text{eff}} = D\left[F_+^2\left(\frac{1 + \cos^2\iota}{2}\right)^2 + F_\times^2 \cos^2\iota\right]^{-1/2} \quad (3.3c)$$

is the *effective distance* of the source. The effective distance of the source is related to the true distance of the source by several geometrical factors that relate the source orientation to the detector orientation. Because the location and orientation of the source are not likely to be known when filtering data from a single detector, it is convenient to combine the geometric factors with the true distance to give a single observable, the effective distance. For an optimally oriented source (one that is directly overhead and is orbiting in the plane of the sky) the effective distance is equal to the true distance; for sub-optimally-oriented sources the effective distance is greater than the true distance. (The location and distance can be estimated using three or more detectors, but we do not consider this here.)

Equation (3.3a) gives a waveform that is used as a template for a matched filter. Since FINDCHIRP implements the matched filter via a FFT correlation, it is beneficial to write the Fourier transform of the template and implement it directly rather than taking the FFT of the time-domain waveform of Eq. (3.3a). A frequency-domain version of the waveform can be obtained via the stationary phase approximation [6,42,66]. For $f > 0$ one has

$$\tilde{h}(f) = -\left(\frac{5\pi}{24}\right)^{1/2}\left(\frac{G\mathcal{M}}{c^3}\right)\left(\frac{G\mathcal{M}}{c^2 D_{\text{eff}}}\right)\left(\frac{G\mathcal{M}}{c^3}\pi f\right)^{-7/6} e^{-i\Psi(f;M,\mu)} = \left(\frac{1\text{ Mpc}}{D_{\text{eff}}}\right)\mathcal{A}_{1\text{ Mpc}}(M,\mu) f^{-7/6} e^{-i\Psi(f;M,\mu)}, \quad (3.4a)$$

where

$$\mathcal{A}_{1\text{ Mpc}}(M, \mu) = -\left(\frac{5}{24\pi}\right)^{1/2}\left(\frac{GM_\odot/c^2}{1\text{ Mpc}}\right)\left(\frac{\pi GM_\odot}{c^3}\right)^{-1/6}\left(\frac{\mathcal{M}}{M_\odot}\right)^{5/6}, \quad (3.4b)$$

$$\Psi(f; M, \mu) = 2\pi f t_0 - 2\phi_0 - \pi/4 + \frac{3}{128\eta}\left[v^{-5} + \left(\frac{3715}{756} + \frac{55}{9}\eta\right)v^{-3} \\ - 16\pi v^{-2} + \left(\frac{15\,293\,365}{508\,032} + \frac{27\,145}{504}\eta + \frac{3085}{72}\eta^2\right)v^{-1}\right], \quad (3.4c)$$

$$v = \left(\frac{GM}{c^3}\pi f\right)^{1/3} \quad (3.4d)$$

and $\Psi$ has been written to second post-Newtonian order. For $f < 0$ one has $\tilde{h}(f) = \tilde{h}^*(-f)$. This template waveform has been expressed in terms of several factors: (1) An overall distance factor involving the effective distance, $D_{\rm eff}$. For a *template* waveform, we are free to choose this effective distance as convenient, and in the FINDCHIRP code it is chosen to be 1 Mpc. (2) A constant (in frequency) factor $\mathcal{A}_{1\,\rm Mpc}(M, \mu)$, which has dimensions of $({\rm time})^{-1/6}$, that depends only on the total and reduced masses, $M$ and $\mu$, of the particular system. (3) The factor $f^{-7/6}$ which does not depend on the system parameters. And (4) a phasing factor involving the phase $\Psi(f; M, \mu)$ which is both frequency-dependent and dependent on the system's total and reduced masses. We will see below that an efficient implementation of the matched filter will make use of this factorization of the stationary phase template.

In order to construct a waveform template we need to know how long the binary system will radiate gravitation waves in the sensitivity band of LIGO. A true inspiral chirp waveform would last tens of Myr, but the amount of time that the binary system spends radiating gravitational waves with a frequency above some low-frequency cutoff $f_{\rm low}$ is short: the duration of the chirp or *chirp time* from a given frequency $f_{\rm low}$ is given to second post-Newtonian order by Eq. (3.3) of Ref. [43] as

$$T_{\rm chirp} = \frac{5}{256\eta} \frac{GM}{c^3} \left[ v_{\rm low}^{-8} + \left( \frac{743}{252} + \frac{11}{3}\eta \right) v_{\rm low}^{-6} - \frac{32\pi}{5} v_{\rm low}^{-5} \right.$$
$$\left. + \left( \frac{3\,058\,673}{508\,032} + \frac{5429}{504}\eta + \frac{617}{72}\eta^2 \right) v_{\rm low}^{-4} \right] \quad (3.5a)$$

where

$$v_{\rm low} = \left( \frac{GM}{c^3}\pi f_{\rm low} \right)^{1/3}. \quad (3.5b)$$

For the Initial and Enhanced LIGO detectors, FINDCHIRP uses $f_{\rm low} = 40$ Hz. Higher-mass systems coalesce much more quickly (from a given $f_{\rm low}$) than lower mass systems. A search for low-mass systems, such as primordial black holes of mass $0.1 M_\odot$, can require very long waveform templates (of the order of tens of minutes) which can result in a significant computational burden.

There is also a high-frequency cutoff for the inspiral waveform. Physically, at some high frequency a binary system will terminate its secular inspiral and the orbit will decay on a dynamical time scale, though identifying the precise frequency is difficult except in the extreme mass ratio limit $\eta \to 0$. In this limit, that of a test mass orbiting a Schwarzschild black hole, the frequency is known as the innermost stable circular orbit or ISCO. The ISCO gravitational-wave frequency is

$$f_{\rm isco} = \frac{c^3}{6\sqrt{6}\pi GM}. \quad (3.6)$$

However, before reaching this frequency, the binary components will be orbiting with sufficiently high orbital velocities that the higher-order corrections to the post-Newtonian waveform will become significant. We regard Eq. (3.6) as an upper limit on the frequency that can be regarded as representing an "inspiral" waveform—not as the frequency to which we can trust our inspiral waveform templates. With this caveat, we nevertheless use this as a high-frequency cutoff for the inspiral template waveforms (should this frequency be less than the Nyquist frequency $f_{\rm Nyquist} = 1/2\Delta t$ of the data, where $\Delta t$ is the sample rate). For the lowest mass binary systems (binary neutron stars or subsolar mass black holes) the post-Newtonian template waveforms are reliable within the sensitive band of Initial LIGO so the precise choice of the high-frequency cutoff is not important [36]. For higher-mass systems (containing black holes) the effects of higher-order corrections to the post-Newtonian waveform can be significant. For frequency-domain templates, this can be addressed by the use of "pseudo post-Newtonian" terms in the waveform phasing and/or different choices of the high-frequency cutoff [44,45]. When using time-domain EOB waveforms tuned to numerical relativity simulations, the upper-frequency cutoff is set by the frequency of the fundamental $l = 2$, $m = 2$ quasinormal ringdown mode [46]. In this case the FINDCHIRP algorithm uses the minimum of this or the Nyquist frequency.

## IV. MATCHED FILTER

The matched filter is the optimal filter for detecting a known waveform in stationary Gaussian noise. Suppose that $n(t)$ is a stationary Gaussian noise process with one-sided power spectral density $S_n(f)$ defined by $\langle \tilde{n}(f)\tilde{n}^*(f')\rangle = \frac{1}{2} S_n(|f|)\delta(f-f')$. Then the matched-filter output of a data stream $s(t)$—which now may contain detector noise alone, $s(t) = n(t)$, or a signal in addition to the noise, $s(t) = n(t) + h(t)$—with a filter template $h_{\rm template}(t)$ is

$$x(t_0) = 2 \int_{-\infty}^{\infty} \frac{\tilde{s}(f)\tilde{h}^*_{\rm template}(f)}{S_n(f)} df$$
$$= 4\Re \int_0^\infty \frac{\tilde{s}(f)[\tilde{h}^*_{\rm template}(f)]_{t_0=0}}{S_n(f)} e^{2\pi i f t_0} df, \quad (4.1)$$

where the signal $h_{\rm template}(t)$ is implicitly taken to depend on a termination time $t_0$ as in (3.3a). Notice that the use of a FFT will allow one to search for all possible termination times $t_0$ efficiently. However, the waveforms described above have additional unknown parameters. These are (i) the amplitude (or effective distance to the source), (ii) the coalescence phase, and (iii) the binary companion masses. The amplitude simply sets a scale for the matched-filter output, and is unimportant for matched-filter templates (these can be normalized).

The "best match" unknown phase $\phi_0$ can be found by maximizing $x(t_0)$ over $\phi_0$. $x(t_0)$ can be written as $x(t_0) = x_{\rm re}(t_0)\cos 2\phi_0 + x_{\rm im}(t_0)\sin 2\phi_0$ where $x_{\rm re,im}(t_0)$ are (respectively) the values of (4.1) with $\phi_0 = 0$, and with $\phi_0 = 0$ and Re $\to$ Im. Setting the derivative with regard to $\phi_0$ to vanish, one finds that at the maximum $x^2(t_0)|_{\hat\phi_0} = x_{\rm re}^2(t_0) + x_{\rm im}^2(t_0)$ at $2\hat\phi_0 = \arg(x_{\rm re} + ix_{\rm im})$. Hence this maximum value is given by the (more computationally efficient) modulus of the complex filter output

$$z(t_0) = x_{\rm re}(t_0) + ix_{\rm im}(t_0)$$
$$= 4\int_0^\infty \frac{\tilde s(f)[\tilde h_{\rm template}^*(f)]_0}{S_n(f)} e^{2\pi i f t_0} df \quad (4.2)$$

where $[\tilde h_{\rm template}^*(f)]_0 = [\tilde h_{\rm template}^*(f)]_{t_0=0,\phi_0=0}$. This template is obtained from Eq. (3.4) by deleting the first two terms on the right-hand side of (3.4c).

To search over all the possible binary companion masses it is necessary to construct a bank of matched-filter templates laid out on the $(m_1, m_2)$ plane sufficiently finely that any true system masses will produce a waveform that is close enough to the nearest template. There are well-known strategies for constructing such a bank [8–11]. For our purposes, we shall simply introduce an index $m = 0, \ldots, N_{\rm T} - 1$ labeling the particular waveform template $h_m(t)$ in the bank of $N_{\rm T}$ waveform templates.

By convention, the waveform templates are constructed for systems with an effective distance of $D_{\rm eff} = 1$ Mpc. To construct a signal-to-noise ratio, a normalization constant for each template is computed

$$\sigma_m^2 = 4\int_0^\infty \frac{|\tilde h_{\rm 1\ Mpc,m}(f)|^2}{S_n(f)} df. \quad (4.3)$$

The quantity $\sigma_m$ is a measure of the sensitivity of the instrument. For $s(t)$ that is purely stationary and Gaussian noise, $\langle \Re z_m(t)\rangle = \langle \Im z_m(t)\rangle = 0$ and one finds that $\langle z_m^2(t)\rangle = \langle z_m^2(t)\rangle = \sigma_m^2$, while for a detector output that includes a signal at distance $D_{\rm eff}$, $s(t) = n(t) + (D_{\rm eff}/1\ {\rm Mpc})^{-1} h_{\rm 1\ Mpc,m}(t)$, $\langle z_m(t_0)\rangle = 1\,{\rm Mpc}\,\sigma_m^2/D_{\rm eff}$. Thus the quantity

$$\rho_m(t) = \frac{|z_m(t)|}{\sigma_m} \quad (4.4)$$

is the *amplitude* signal-to-noise ratio of the (quadrature) matched filter. Note that while this is called a signal-to-noise ratio, in the absence of a signal the expectation value is not unity: $\langle \rho_m^2 \rangle = 2$. It is highly unlikely to obtain $\rho_m \gg 1$ for purely stationary and Gaussian noise so a detection strategy usually involves setting a lower threshold on $\rho_m$ to identify event candidates. For such candidates, a biased estimate of the effective distance to the candidate system is $\hat D_{\rm eff} = (\sigma_m/\rho_m)\,{\rm Mpc}$.

The goal of the FINDCHIRP algorithm is largely to construct the quantity $\rho_m(t)$ and to identify the values of the parameters $t_0$, $\phi_0$, and $m$ that maximize it.

## V. DETECTOR OUTPUT AND CALIBRATION

LIGO records several interferometer channels. The *gravitational-wave channel* (the primary channel for searching for gravitational waves) is formed from the output of a photo-diode at the antisymmetric (or "dark") port of the interferometer [47]. This output is used as an error signal for a feedback loop that is needed to keep various optical cavities in the interferometer in resonance or "in-lock." Hence it is often called the error signal $e(t)$. The error signal is not an exact measure of the differential arm displacements of the interferometer so it does not correspond to the gravitational-wave strain. Rather it is part of a linear feedback loop that controls the position of the interferometer mirrors. A gravitational-wave strain-equivalent output, called $s(t)$ above, can be obtained from the error signal $e(t)$ via a linear filter. This is called *calibration*. Details on the calibration of the LIGO interferometers can be found in [48,49]. In the frequency-domain, the process of calibration can be thought of as multiplying the error signal by a complex *response function*, $R(f)$

$$\tilde s(f) = R(f)\tilde e(f). \quad (5.1)$$

The FINDCHIRP algorithm can analyze either calibrated data $s(t)$ or it can calibrate the error signal $e(t)$ in the frequency-domain. With time-domain calibrated data now available, it is more convenient to use $s(t)$ so in this paper we will present the algorithm in terms of strain data rather than the error signal.

The detector output is not a continuous signal but rather a time series of samples of $s(t)$ taken with a sample rate of $1/\Delta t = 16384$ Hz where $\Delta t$ is the sampling interval. Thus, rather than $s(t)$, the input to FINDCHIRP is a discretely sampled set of values $s[j] = s(t_{\rm start} + j\Delta t)$ for some large number of points. The start of the data sample is at time $t_{\rm start}$. Data from the detector is divided into *science segments* which are time epochs when the instrument was in-lock and exhibiting normal behavior. However, for practical computational reasons, these science segments are not normally processed as a whole, but are divided into smaller amounts. In this paper we shall call the amount of data processed a *data block* of duration $T_{\rm block}$. The data block must be long enough to form a reliable noise power spectral estimate (see below), but not so long as to exhaust a computer's memory or to encompass significant nonstationary changes in the detector noise.

The number of points in a data block is further subdivided into $N_{\rm S}$ overlapping *data segments* or just *segments* (not to be confused with the science segments described above) of duration $T$. The duration of the segment is always an integer multiple of the sample rate $\Delta t$, so the number of points in a segment $N = T/\Delta t$ is an integer. These segments are used to construct an average noise power spectrum and to perform the matched filtering. The segments are overlapped so that the first segment consists of the points $s[j]$ for $j = 0, \ldots, N-1$, the second consists of the points $j = \Delta, \ldots, \Delta + N - 1$ where $\Delta$ is known as the *stride*, and so on until the last segment which consists of the points $j = (N_S - 1)\Delta, \ldots, (N_S - 1)\Delta + N - 1$. Note that

$$T_{\text{block}} = [(N_S - 1)\Delta + N]\Delta t. \quad (5.2)$$

All of the LSC's searches choose to overlap the segments by 50% so that the stride is $\Delta = N/2$ (and $N$ is always even) and hence there are $N_S = 2(T_{\text{block}}/T) - 1$ segments. The values of $T_{\text{block}}$, $T$, $\Delta$, and $N_S$ must be commensurate so that these relations hold. Values typically used in the LIGO and/or Virgo analyses for low-mass binaries are $T_{\text{block}} = 2048\,\text{s}$, $N = 256\,\text{s}/\Delta t$, $N_S = 15$, and $\Delta = 128\,\text{s}/\Delta t$.

The FINDCHIRP algorithm (4.2) implements the matched filter by an FFT correlation. Thus the discrete Fourier transforms of the individual data segments, $n$,

$$\tilde{s}_n[k] = \Delta t \sum_{j=0}^{N-1} s[j + n\Delta] e^{-2\pi ijk/N} \quad (5.3)$$

for $n = 0, \ldots, N_S - 1$ are constructed via an FFT. Here $k$ is a frequency index that runs from 0 to $N - 1$. The $k = 0$ component represents the DC component ($f = 0$) which is purely real, the components $0 < k \leq \lfloor (N-1)/2 \rfloor$ are all positive-frequency components corresponding to frequencies $k\Delta f$ where $\Delta f = 1/(N\Delta t)$, and the components $\lfloor N/2 \rfloor < k < N$ are all negative-frequency components corresponding to frequencies $(k - N)\Delta f$. If $N$ is even (as it always is for the FINDCHIRP algorithm) then there is also a purely real Nyquist-frequency component in bin $k = N/2$ corresponding to the Nyquist frequency $\pm N\Delta f/2 = \pm 1/(2\Delta t)$. Recall $\lfloor a \rfloor$ is the greatest integer less than or equal to $a$. Note that because the data is real, the discrete Fourier transform of it satisfies $\tilde{s}_n^*[k] = \tilde{s}_n[N - k]$. Thus, the FINDCHIRP algorithm only stores the frequency components $k = 0, \ldots, \lfloor N/2 \rfloor$. These can be efficiently computed using a *real-to-half-complex forward FFT* [50].

If the error signal rather than detector strain is analyzed, we simply replace $s$ with $e$ in Eq. (5.3). Then the detector strain for segment $n$ can be computed by calibrating the error signal

$$\tilde{s}_n[k] = R[k]\tilde{e}_n[k] \text{ for } k = 0, \ldots, \lfloor N/2 \rfloor, \quad (5.4)$$

where $R[k] = R(k\Delta f)$ is the complex response function. As before, since $\tilde{s}_n[k]$ must be the Fourier transform of some real time series, only the frequency components $k = 0, \ldots, \lfloor N/2 \rfloor$ need to be computed.

LIGO is sensitive to strains that are smaller than $\sim 10^{-20}$, while the error signal is designed to have typical values much closer to unity. Often the FINDCHIRP algorithm will require quantities that are essentially squares of the measured strain (e.g., the power spectrum described in the next section). To avoid floating-point over- or under-flow problems, the strain can simply be rescaled by a dynamic-range factor $\kappa$. If strain data is input, it is immediately scaled by the factor $\kappa$, so $\kappa s[j]$ is used rather than $s[j]$. If the error signal is input, the dynamic-range factor is applied to the response function instead, so that $\kappa R$ is used rather than $R$. Choosing a value of $\kappa \sim 10^{20}$ will keep all quantities within representable single-precision IEEE 781 floating-point numbers. It is important to keep track of the factor $\kappa$ to make sure it cancels out in all of the results. Essentially this is achieved by multiplying all quantities with "units" of strain by the factor $\kappa$ within the implementation of the FINDCHIRP algorithm. Thus, in addition to the strain data or response function, the signal template must also be scaled by $\kappa$. On conventional CPUs, storing quantities in single-precision reduces the performance cost of moving quantities to and from memory (which can be dominant). In addition, some CPUs and most graphics processing units (GPUs) perform single-precision arithmetic at least twice as fast as double-precision arithmetic. Therefore it is advantageous to use floating-point (single-precision) operations for the FINDCHIRP algorithm.

If frequency-domain calibration of the error signal is desired (e.g., if the effects of calibration error are being investigated), the following replacements need to be made in the formulas in this paper: $\tilde{s}[k] \to R[k]\tilde{e}[k]$, $S[k] \to |R[k]|^2 S_e[k]$, and $Q \to |R[k]|^{-2} Q_e[k]$ where $S_e[k]$ and $Q_e[k]$ are the power spectral density and inverse truncated power spectral density computed from the error signal rather than the calibrated strain data (see Secs. VI and VII).

## VI. AVERAGE POWER SPECTRUM

Part of the matched filter involves weighting the data by the inverse of the detector's power spectral density. The detector's power spectrum must be obtained from the detector output. The most common method of power spectral estimation is Welch's method. Welch's method [51] for obtaining the average power spectrum $S$ of the data is

$$\kappa^2 S[k] = \frac{1}{N_S} \sum_{n=0}^{N_S - 1} \kappa^2 P_n[k]. \quad (6.1)$$

Here

$$\kappa^2 P_n[k] = \frac{2\Delta f}{W} \left| \Delta t \sum_{j=0}^{N-1} \kappa s_n[j] w[j] e^{-2\pi ijk/N} \right|^2 \quad (6.2\text{a})$$

is a normalized periodogram for a single segment $n$ which is the modulus-squared of the discrete Fourier transform of *windowed* data. The data window is given by $w[j]$ and $W$ is a normalization constant

$$W = \frac{1}{N} \sum_{j=0}^{N-1} w^2[j]. \quad (6.2\text{b})$$

FINDCHIRP allows a variety of possible windows, but a Hann window (see, e.g., [52]) is the default choice used by FINDCHIRP. We call this average power spectrum the *mean average power spectrum*. If uncalibrated data $e_n[j]$ is being used rather than $s_n[j]$, then the power spectrum of the detector strain-equivalent noise is related to the power spectrum of the uncalibrated data by $S[k] = |R[k]|^2 S_e[k]$.

The problem with using Welch's method for power spectral estimation is that for detector noise containing significant excursions from "normal" behavior (due to instrumental glitches or unexpectedly strong gravitational-wave signals), the mean used in Eq. (6.1) can be significantly biased by the excursion. An alternative that is pursued in the FINDCHIRP algorithm is to replace the mean in Eq. (6.1) by a median, which is a more robust estimator of the average power spectrum

$$\kappa^2 S[k] = \alpha^{-1} \operatorname{median}\{\kappa^2 P_0[k], \kappa^2 P_1[k], \ldots, \kappa^2 P_{N_S-1}[k]\},$$
(6.3a)

where $\alpha$ is a required correction factor. When $\alpha = 1$, the expectation value of the median is not equal to the expectation value of the mean in the case of Gaussian noise; hence the factor $\alpha$ is introduced to ensure that the same power spectrum results for Gaussian noise. In Ref. [53] and in Appendix B it is shown that if the set $\{P_0[k], P_1[k], \ldots, P_{N_S-1}[k]\}$ are independent exponentially-distributed random variables (as expected for Gaussian noise) then

$$\alpha = \sum_{n=1}^{N_S} \frac{(-1)^{n+1}}{n} \quad (\text{odd } N_S)$$
(6.3b)

is the correction factor. We call this median estimate of the average power spectrum, corrected by the factor $\alpha$, the *median average spectrum*.

Unfortunately this result is not exactly correct either. Because the segments used to form the individual sample values $P_n[k]$ of the power at a given frequency are somewhat overlapping (unless $\Delta \gtrsim N$), they are not independent random variables, as was assumed in Appendix B. (This is somewhat mitigated by the windowing of the segments of data.) Although the effect is not large, and simply amounts to a slight scaling of what is meant by signal-to-noise ratio, we are led to propose a variant of the median method in which the $n = 0, \ldots, N_S - 1$ overlapping segments are divided into *even segments* (for which $n$ is even) and the *odd segments* (for which $n$ is odd). If the stride is $\Delta \geq N/2$ then no two even segments will depend on the same data so the even segments will be independent; similarly the odd segments will be independent. The average power spectrum can be estimated by taking the weighted mean of the median power spectrum of the $(N_S + 1)/2$ even segments and the median power spectrum of the $(N_S - 1)/2$ odd segments, each of which are corrected by a factor $\alpha$ appropriate for the sample median with, respectively, $(N_S \pm 1)/2$ samples. We call this the *median-mean average spectrum*. Like the median spectrum it is not overly sensitive to a single glitch (or strong gravitational-wave signal).

The FINDCHIRP algorithm can compute the mean average spectrum, the median average spectrum, or the median-mean average spectrum. Traditionally the median average spectrum has been used though we expect that the median-mean average spectrum will be adopted in the future.

## VII. DISCRETE MATCHED FILTER

The discretized version of Eq. (4.2) is simply

$$z_{n,m}[j] = 4\Delta f \sum_{k=1}^{\lfloor (N-1)/2 \rfloor} \frac{\kappa \tilde{s}_n[k] \kappa \tilde{h}^*_{1\text{ Mpc},m}[k]}{\kappa^2 S[k]} e^{2\pi i j k/N}, \quad (7.1)$$

where in $\tilde{h}$ both $t_0$ and $\phi_0$ are set to zero. Element $j$ of $z_{n,m}[j]$ corresponds to the matched-filter output (4.2) for time $t_0 = t_{\text{start}} + (n\Delta + j)\Delta t$ where $t_{\text{start}}$ is the start time of the block of data analyzed and $\Delta$ is the stride. Note that the sum is over the positive frequencies only, and DC and Nyquist frequencies are excluded. (The interferometer is AC-coupled so it has no sensitivity at the DC component; similarly, the instrument has very little sensitivity at the Nyquist-frequency so rejecting this frequency bin has very little effect.) This inverse Fourier transform can be performed by the *complex* reverse DFT (as opposed to a half-complex-to-real reverse DFT) of the quantity

$$\tilde{z}_{n,m}[k]\Delta f$$
$$= \begin{cases} 0 & k < k_{\text{low}} \\ 4\Delta f \dfrac{\kappa \tilde{s}_n[k] \kappa \tilde{h}^*_{1\text{ Mpc},m}[k]}{\kappa^2 S[k]} & k_{\text{low}} \leq k \leq \lfloor (N-1)/2 \rfloor \\ 0 & \lfloor (N-1)/2 \rfloor < k < N. \end{cases}$$
(7.2)

That is to say, the DC, Nyquist, and negative frequency components are all set to zero, as are all frequencies below some low-frequency cutoff $f_{\text{low}} = k_{\text{low}} \Delta f$ (set to some frequency lower than the detector's sensitive band). The low-frequency cutoff limits the time duration of the inspiral template as described below.

Our task is to obtain an efficient computation of the factors making up $\tilde{z}_{n,m}[k]$. Note that there needs to be one reverse FFT performed per segment per template. It is desirable that this (unavoidable) computational cost dominates the evaluation of the matched filter, so we wish to make the computation cost of the calculation of $\tilde{z}_{n,m}[k]$ for all $k$ and fixed $n$, $m$ to be less than the computation cost of a FFT. We will consider this in the next section.

One subtlety in the construction of the matched filter is the issue of filter wraparound. The matched filter of Eq. (7.1) can be thought of as digital correlation of a filter $h_{1\text{ Mpc},m}[j]$ with some suitably over-whitened data stream (the data divided by the noise power spectrum). To simplify the discussion, first assume $S[k] = 1$. Although $\tilde{h}_{1\text{ Mpc},m}[k]$ is generated in the frequency domain via the stationary phase approximation, we can imagine that it came from a time-domain signal $h_{1\text{ Mpc},m}[j]$ of duration $T_{\text{chirp},m}$ that is given by Eq. (3.5a) for the low-frequency cutoff $f_{\text{low}}$. By convention of template generation, the time of coalescence corresponds to $j = 0$. Thus, the entire chirp waveform is nonzero only from $t = t_0 - T_{\text{chirp},m}$ to $t = t_0$. Because the discrete Fourier transform presumes that the data is periodic, this is represented by having the chirp begin at point $j = N - T_{\text{chirp},m}/\Delta t$ and end at point $j = N - 1$. Thus $h_{1\text{ Mpc},m}[j] = 0$ for $j = 0, \ldots, N - 1 - T_{\text{chirp},m}/\Delta t$. The correlation of $h_{1\text{ Mpc},m}[j]$ with the interferometer data will involve multiplying the $T_{\text{chirp},m}/\Delta t$ points of data *before* a given time with the $T_{\text{chirp},m}/\Delta t$ points of the chirp. When this is performed by the FFT correlation, this means that the first $T_{\text{chirp},m}/\Delta t$ points of the matched-filter output involve data at times before the start of the segment, which are interpreted as the data values at the end of the segment (since the FFT assumes that the data is periodic). Hence the first $T_{\text{chirp},m}/\Delta t$ points are of the correlation are invalid and must be discarded. That is, of the $N$ points of $z_{n,m}[j]$ in Equation (7.1), only the points $j = T_{\text{chirp},m}/\Delta t, \ldots, N - 1$ are valid. Recall that the analysis segments of data are overlapped by an amount $N - \Delta$: this is to ensure that the matched-filter output is continuous (except at the very beginning of a data block). That is, only points $j = T_{\text{chirp},m}/\Delta t, \ldots, N - 1$ of $z_{0,m}$ are valid and only points $j = T_{\text{chirp},m}/\Delta t, \ldots, N - 1$ of $z_{1,m}$ are valid, but points $j = \Delta, \ldots, N - 1$ of $z_{0,m}[j]$ correspond to points $j = 0, \ldots, N - \Delta - 1$ of $z_{1,m}[j]$, and these can be used instead. Therefore FINDCHIRP must ensure that the amount that the data segments overlap, $N - \Delta$ points, is greater than or equal to the duration, $T_{\text{chirp},m}/\Delta t$ points, of the filter: $T_{\text{chirp},m}/\Delta t \leq N - \Delta$.

The quantity that needs to be computed in Eq. (7.1) is more than just a correlation of the data $s_n[j]$ with the filter $h_{1\text{ Mpc},m}[j]$: it also involves a convolution of the data with the response function (if we are performing frequency-domain calibration) and the inverse of the power spectrum. The interferometer has a relatively short impulse response, so this convolution will only corrupt a short amount of data (though now at the end as well as at the beginning of a analysis segment). However, the inverse of the power spectrum has many very narrow line features that act as sharp notch filters when applied to the data. These filters have an impulse response that is as long as the reciprocal of the resolution of the frequency series, which is set by the amount of data used to compute the periodograms that are used to obtain the average spectrum. Since this is the same duration as the analysis segment duration, the convolution of the data with the inverse power spectrum corrupts the *entire* matched-filter output.

To resolve this problem we apply a procedure to coarse-grain the inverse power spectrum called *inverse spectrum truncation*. Our goal is to limit the amount of the matched filter that is corrupted due to the convolution of the data with the inverse power spectrum. To do this we will begin with the time-domain version of the frequency-domain quantity $S^{-1}[k]$, truncate it so that it has finite duration, and then find the frequency-domain quantity $Q[k]$ corresponding to this truncated time-domain filter. Note that $S^{-1}[k]$ is real and non-negative, and we want $Q[k]$ to share these properties. First, construct the time-domain quantity

$$\kappa^{-1}q[j] = \Delta f \sum_{k=0}^{N-1} \sqrt{1/(\kappa^2 S[k])}\, e^{2\pi i j k/N}, \qquad (7.3a)$$

which can be done via a half-complex-to-real reverse FFT. Since $S[k]$ is real and symmetric ($S[k] = S[N-k]$), $q[j]$ will be real and symmetric (so that $q[j] = q[N-j]$). This quantity will be nonzero for all $N$ points, though strongly-peaked near $j = 0$ and $j = N - 1$. Now create a truncated quantity $q_T[j]$ with a total duration of $T_{\text{spec}}$ ($T_{\text{spec}}/2$ at the beginning and $T_{\text{spec}}/2$ at the end)

$$\kappa^{-1}q_T[j] = \begin{cases} \kappa^{-1}q[j] & 0 \leq j < T_{\text{spec}}/2\Delta t \\ 0 & T_{\text{spec}}/2\Delta t \leq j < N - T_{\text{spec}}/2\Delta t \\ \kappa^{-1}q[j] & N - T_{\text{spec}}/2\Delta t \leq j < N. \end{cases}$$
(7.3b)

Since $q_T$ is real and symmetric, the discrete Fourier transform of $q_T$ will also be real and symmetric, though not necessarily positive. Therefore we construct

$$\kappa^{-2}Q[k] = \left| \Delta t \sum_{j=0}^{N-1} \kappa^{-1} q_T[j] e^{-2\pi i j k/N} \right|^2 \qquad (7.3c)$$

This quantity is real, positive, and symmetric, as desired. Multiplying the data by $Q[k]$ in the frequency-domain is equivalent to convolving the data with $q_T[j]$ in the time-domain *twice*, which will have the effect of corrupting a duration of $T_{\text{spec}}$ of the matched filter $z_{n,m}[j]$ at the beginning and a duration of $T_{\text{spec}}$ at the end of the data segment. This is in addition to the duration $T_{\text{chirp},m}$ that is corrupted at the beginning of the data segment due to the correlation with the filter $h_{1\text{ Mpc},m}[j]$. Thus the total duration that is corrupted is $2T_{\text{spec}} + T_{\text{chirp},m}$, and this must be less than the time that adjacent segments overlap. The net effect of the inverse spectrum truncation is to smear out sharp spectral features and to decrease the resolution of the inverse power spectrum weighting.

For simplicity, we normally choose a 50% overlap (so that $\Delta = N/2$). Of each data segment the middle half with $j = N/4, \ldots, 3N/4 - 1$ is assumed to be valid matched-filter output. Therefore, the inverse truncation duration $T_{\text{spec}}$ and the maximum filter duration $T_{\text{chirp},m}$ must satisfy $T_{\text{spec}} + T_{\text{chirp},m} \leq T/4$ since a time $T_{\text{spec}} + T_{\text{chirp},m}$ is corrupted at the beginning of the data segment.

## VIII. WAVEFORM DECOMPOSITION

Our goal is now to construct the quantity

$$\tilde{z}_{n,m}[k] = 4\kappa^{-2}Q[k]\kappa\tilde{s}_n[k]\kappa\tilde{h}^*_{1\text{ Mpc},m}[k] \quad (8.1)$$

as efficiently as possible. This quantity must be computed for every segment $n$, every template $m$, and every frequency bin in the range $k = k_{\text{low}}, \ldots, k_{\text{high},m} - 1$ where $k_{\text{low}} = \lfloor f_{\text{low}}/\Delta f \rfloor$ and $k_{\text{high},m}$ is the high-frequency cutoff of the waveform template, which is given by the minimum of the ISCO frequency of Eq. (3.6) and the Nyquist-frequency

$$k_{\text{high},m} = \min\{\lfloor f_{\text{isco}}/\Delta f \rfloor, \lfloor (N+1)/2 \rfloor\} \quad (8.2)$$

(recall that the ISCO frequency depends on the binary system's total mass so it is template-dependent). We can factorize $\tilde{z}_{n,m}[k]$ as follows:

$$\tilde{z}_{n,m}[k] = (\Delta f)^{-1} A_{1\text{ Mpc},m} F_n[k] G_m[k] \quad (8.3)$$

where $A_{1\text{ Mpc},m}$ is a *template normalization* (it needs to be computed once per template but does not depend on $k$), $G_m[k] = \exp(i\Psi_m[k])$ with $\Psi_m[k]$ is a *template phase* which must be computed at all values of $k$ for every template (but does not depend on the data segment), and $F_n[k]$ is the FINDCHIRP *data segment* that must be computed for all values of $k$ for each data segment (but does not depend on the template). As before, both $t_0$ and $\phi_0$ are set to zero in $\Psi$. FINDCHIRP first computes and stores the quantities $F_n[k]$ for all data segments. Then, for each template $m$ in the bank, the phasing $\Psi_m[k]$ is computed once and then applied to all of the data segments (thereby marginalizing the cost of the template generation).

To facilitate the factorization, we rewrite Eq. (3.4a) with $t_0 = \phi_0 = 0$ in the discrete form

$$\kappa\tilde{h}_{1\text{ Mpc},m}[k] = (\Delta f)^{-1} A_{1\text{ Mpc},m} k^{-7/6} \exp(-i\Psi_m[k]) \quad (8.4)$$

with

$$A_{1\text{ Mpc},m} = -\kappa\left(\frac{5}{24\pi}\right)^{1/2}\left(\frac{GM_\odot/c^2}{1\text{ Mpc}}\right)\left(\frac{GM_\odot}{c^3}\pi\Delta f\right)^{-1/6}\left(\frac{\mathcal{M}}{M_\odot}\right)^{5/6}, \quad (8.5a)$$

$$\Psi_m[k] = -\pi/4 + \frac{3}{128\eta}\left[v_m^{-5}[k] + \left(\frac{3715}{756} + \frac{55}{9}\eta\right)v_m^{-3}[k] - 16\pi v_m^{-2}[k]\right.$$
$$\left. + \left(\frac{15\,293\,365}{508\,032} + \frac{27\,145}{504}\eta + \frac{3085}{72}\eta^2\right)v_m^{-1}[k]\right], \quad (8.5b)$$

$$v_m[k] = \left(\frac{GM_\odot}{c^3}\pi\Delta f\right)^{1/3}\left(\frac{M}{M_\odot}\right)^{1/3} k^{1/3}. \quad (8.5c)$$

The dependence on the data segment is wholly contained in the template-independent quantity $F_n[k]$ which is

$$F_n[k] = 4\kappa^{-2} Q[k]\kappa\tilde{s}_n[k] k^{-7/6}. \quad (8.6)$$

As mentioned earlier, FINDCHIRP computes and stores $F_n[k]$ for all segments only once, and then reuses these precomputed spectra in forming $\tilde{z}_{n,m}[k]$ according to Eq. (8.3). The $k$-dependence on the template is wholly contained in the data-segment-independent quantity $G_m[k]$ which is

$$G_m[k] = \begin{cases} \exp(i\Psi_m[k]) & k_{\text{low}} \leq k < k_{\text{high},m} \\ 0 & \text{otherwise} \end{cases}. \quad (8.7)$$

This quantity is known as the FINDCHIRP *template*.

The value of $\sigma_m$ is also needed in order to normalize $z_{n,m}[j]$ to compute signal-to-noise $\rho$. It is

$$\sigma_m^2 = 4\Delta f \sum_{k=k_{\text{low}}}^{k_{\text{high},m}-1} \kappa^{-2}Q[k]|\kappa\tilde{h}_{1\text{ Mpc},m}[k]|^2$$
$$= A_{1\text{ Mpc},m}^2 \mathsf{s}^2[k_{\text{high},m}], \quad (8.8)$$

where

$$\mathsf{s}^2[k_{\text{high},m}] = \frac{4}{\Delta f}\sum_{k=k_{\text{low}}}^{k_{\text{high},m}-1} \kappa^{-2}Q[k] k^{-7/3} \quad (8.9)$$

needs to be computed only once per data block (i.e., only once per power spectrum)—it does not depend on the particular segment within a block or on the particular template in the bank, except in the high-frequency cutoff of the template (if it is less than the Nyquist-frequency). To account for this minimal dependence on the template, the quantity $\mathsf{s}[k_{\text{high}}]$ is precomputed for all values of $k_{\text{high}}$.

The division of the matched filter into the data-segment-only quantity $F_n[k]$ and the template-only quantity $G_m[k]$ means that FINDCHIRP can efficiently compute the matched filter, or, rather, a quantity that is proportional to it

$$\zeta_{m,n}[j] = \sum_{k=0}^{N-1} F_n[k] G_m[k] e^{2\pi i j k/N}. \quad (8.10)$$

Notice that $\zeta_{m,n}[j]$, which is a complex quantity, can be computed using a simple unnormalized reverse FFT of the quantity $F_n[k]G_m[k]$. FINDCHIRP computes and stores $F_n[k]$ for each of the $N_S$ segments in the data block and then, for each template $m$ in the bank, $G_m[k]$ is computed and used to filter each of the $N_S$ segments. This means that ALLEN *et al.*PHYSICAL REVIEW D **85,** 122006 (2012)for each data segment and template the computational cost is essentially limited to $\sim N$ complex multiplications plus one complex reverse FFT.

The quantities $\zeta_{m,n}[j]$ and $z_{m,n}[j]$ are simply related by a normalization factor

$$z_{m,n}[j] = A_{1 \text{ Mpc},m} \zeta_{m,n}[j]. \tag{8.11}$$

Furthermore, the signal-to-noise ratio is related to $\zeta_{m,n}[j]$ via

$$\rho_{m,n}^2[j] = |\zeta_{m,n}[j]|^2/\mathsf{s}^2[k_{\text{high},m}]. \tag{8.12}$$

Rather than applying this normalization to construct the signal-to-noise ratio, FINDCHIRP instead scales the desired signal-to-noise ratio threshold $\rho_\star$ to obtain a normalized threshold

$$\zeta_\star^2 = \mathsf{s}^2[k_{\text{high},m}]\rho_\star^2 \tag{8.13}$$

which can be directly compared to the values $|\zeta_{m,n}[j]|^2$ to determine if there is a candidate event (when $|\zeta_{m,n}[j]|^2 > \zeta_\star^2$). When an event candidate is located, the value of the signal-to-noise ratio can then be recovered for that event along with an estimate of the termination time,

$$\hat{t}_0 = t_{\text{start}} + (n\Delta + j_{\text{peak}})\Delta t, \tag{8.14a}$$

where $j_{\text{peak}}$ is the point at which $|\zeta_{m,n}[j]|$ is peaked; the effective distance of the candidate,

$$\hat{D}_{\text{eff}} = \frac{|A_{1 \text{ Mpc},m}|\mathsf{s}^2[k_{\text{high},m}]}{|\zeta_{m,n}[j_{\text{peak}}]|} \text{ Mpc}; \tag{8.14b}$$

and the termination phase of the candidate,

$$2\hat{\phi}_0 = \arg(z_{mn}[j_{\text{peak}}]) = \arg(-\zeta_{m,n}[j_{\text{peak}}]). \tag{8.14c}$$

The relative sign in the argument arises because $z_{m,n}$ and $\zeta_{m,n}$ have opposite sign; this is because $A_{1\text{Mpc},m}$, which appears in Eq. (8.11) is negative.

## IX. THE CHI-SQUARED VETO

The FINDCHIRP algorithm employs the chi-squared discriminator of Ref. [13] to distinguish between plausible signal candidates and common types of noise artifacts. This method is a type of time-frequency decomposition that ensures that the matched-filter output has the expected accumulation in various frequency bands. (Noise artifacts tend to excite the matched filter at the high-frequency or the low-frequency, but seldom produce the same spectrum as an inspiral.)

For data consisting of pure Gaussian noise, the real and imaginary parts of $\zeta_{m,n}[j]$ (for a given value of $j$) are independent Gaussian random variables with zero mean and variance $\mathsf{s}^2[k_{\text{high},m}]$. If there is a signal present at an effective distance $D_{\text{eff}}$ then $\langle\text{Re}\zeta_{m,n}[j]\rangle = -A_{1\text{Mpc},m}\mathsf{s}^2[k_{\text{high},m}](1\text{Mpc}/D_{\text{eff}})\cos 2\phi_0$ and $\langle\text{Im}\zeta_{m,n}[j]\rangle = -A_{1 \text{ Mpc},m}\mathsf{s}^2[k_{\text{high},m}](1 \text{ Mpc}/D_{\text{eff}})\sin 2\phi_0$ (at the termination time, where $\phi_0$ is the termination phase).

Now consider the contribution to $\zeta_{m,n}[j]$ coming from various frequency sub-bands

$$\zeta_{\ell,m,n}[j] = \sum_{k=k_{(\ell-1)}}^{(k_\ell-1)} F_n[k]G_m[k]e^{2\pi ijk/N} \tag{9.1}$$

for $\ell = 1,\ldots, p$. The $p$ sub-bands are defined by the frequency boundaries $\{k_0 = k_{\text{low}}, k_1, \ldots, k_p = k_{\text{high},m}\}$, which are chosen so that a true signal will contribute an equal amount to the total matched filter from each frequency band. This means that the values of $k_\ell$ must be chosen so that

$$\frac{4}{\Delta f} \sum_{k=k_{(\ell-1)}}^{(k_\ell-1)} \kappa^{-2} Q[k] k^{-7/3} = \frac{1}{p}\mathsf{s}^2[k_{\text{high},m}]. \tag{9.2}$$

With this choice of bands and for pure Gaussian noise, the real and imaginary parts of $\zeta_{\ell,m,n}[j]$ will be independent Gaussian random variables with zero mean and variance $\mathsf{s}^2[k_{\text{high},m}]/p$. Furthermore, the real and imaginary parts of $\zeta_{\ell,m,n}[j]$ and $\zeta_{\ell',m,n}[j]$ with $\ell \neq \ell'$ will be independent since $\zeta_{\ell,m,n}[j]$ and $\zeta_{\ell',m,n}[j]$ are constructed from disjoint bands. Also note that

$$\zeta_{m,n}[j] = \sum_{\ell=1}^{p} \zeta_{\ell,m,n}[j]. \tag{9.3}$$

The chi-squared statistic is now constructed from $\zeta_{\ell,m,n}[j]$ as follows:

$$\chi_{m,n}^2[j] = \sum_{\ell=1}^{p} \frac{|\zeta_{\ell,m,n}[j] - \zeta_{m,n}[j]/p|^2}{\mathsf{s}^2[k_{\text{high},m}]/p}, \tag{9.4}$$

For pure Gaussian noise, $\chi^2$ is chi-squared distributed with $\nu = 2p - 2$ degrees of freedom. That $\nu = 2p - 2$ rather than $\nu = 2p$ results from the fact that the sample mean $\zeta_{m,n}[j]/p$ is subtracted from each of values of $\zeta_{\ell,m,n}[j]$ in the sum. However, this subtraction guarantees that, in the presence of a signal that exactly matches the template $h_{1 \text{ Mpc},m}$ (up to an arbitrary amplitude factor and a coalescence phase), the value of $\chi^2$ is unchanged. Thus, $\chi^2$ is chi-squared distributed with $\nu = 2p - 2$ degrees of freedom in Gaussian noise with or without the presence of an exactly matched signal.

If there is a small mismatch between a signal present in the data and the template, which would be expected since the templates are spaced on a grid and are expected to provide a close match but not a perfect match to a true signal, then $\chi^2$ will acquire a small noncentral parameter. This is because the mismatched signal may not shift the mean value of the real parts of $\{\zeta_{\ell,m,n}\}$ by the same amount (for each $\ell$), and similarly the mean values of the imaginary parts of $\{\zeta_{\ell,m,n}\}$ may not be shifted by the same amounts. The effect on the chi-squared distribution is to introduce a noncentral parameter that is no larger than

122006-10

$\lambda_{\max} = 2\delta\sigma_m^2/D_{\text{eff}}^2$ where $D_{\text{eff}}$ is the effective distance of the true signal and $\delta$ is the *mismatch* between the true signal and the template $h_{1\text{ Mpc},m}$, which could be as large as the *maximum mismatch* of the template bank that is used [8].

Even for small values of $\delta$ (3% is a canonical value), a large value of $\chi^2$ can be obtained for gravitational waves from nearby binaries. Therefore, one should not adopt a fixed threshold on $\chi^2$ lest very loud binary inspirals be rejected by the veto. For a noncentral chi-squared distribution with $\nu$ degrees of freedom and a noncentral parameter of $\lambda$, the mean of the distribution is $\nu + \lambda$ while the variance is between 1 and 2 times the mean (the variance equals twice the mean when $\lambda = 0$ and the variance equals the mean for $\lambda \gg \nu$). Thus it is useful to adopt a threshold on the quantity $\chi^2/(\nu + \lambda)$, which would be expected to be on the order of unity even for very large signals. The FINDCHIRP algorithm adopts a threshold on the related quantity

$$\Xi_{m,n}[j] = \frac{\chi^2_{m,n}[j]}{p + \delta\rho^2_{m,n}[j]}. \tag{9.5}$$

Sometimes the quantity $r^2 = \chi^2/p$ is referred to, rather than $\Xi$, but this quantity does not include the effect of the noncentral parameter.

## X. TRIGGER SELECTION

The signal-to-noise ratio threshold is the primary parameter in identifying candidate events or *triggers*. As we have said, the FINDCHIRP algorithm does not directly compute the signal-to-noise ratio, but rather the quantity $\zeta_{m,n}[j]$ given in Eq. (8.10), whose square modulus is then compared to a normalized threshold given by Eq. (8.13). The computational cost of the search is essentially the cost of $O(N)$ complex multiplications plus $O(N \log N)$ operations to perform the reverse FFT of Eq. (8.10), and an additional $O(N)$ operations to form the square modulus of $\zeta_{m,n}[j]$ for all $j$. In practice, the computational cost is dominated by the reverse complex FFT.

Triggers that exceed the signal-to-noise ratio threshold are then subjected to a chi-squared test. However, the construction of $\chi^2_{m,n}[j]$ is much more costly than the construction of $\zeta_{m,n}[j]$ simply because $p$ reverse complex FFTs of the form given by Eq. (9.1) must be performed.[2] The cost of performing a chi-squared test is essentially $p$ times as great as the cost of performing the matched filter. FINDCHIRP will only perform the chi-squared test if a threshold-crossing trigger is found. Therefore, if threshold-crossing triggers are rare then the cost of the chi-squared test is small compared to the cost of the filtering. Early LIGO science runs (S1 and S2) used a "single-stage" pipeline, in which the chi-squared veto is computed as described here. For reasons of computational cost, later LIGO and/or Virgo analyses used a "two-stage" pipeline in which triggers are required to be coincident between at least two different detectors before the chi-squared is computed. This is achieved by disabling the chi-squared test on the first pass of trigger generation on individual detectors and then only applying it on the triggers that survive the coincidence criteria. The chi-squared computation is easily parallelizable and a GPU can be used to perform the calculation. Simple GPU implementations of the chi-squared veto have been shown to reduce the cost of performing the chi-squared by a factor of $\sim 20$ [54,55]. Given this reduced cost, future LSC analyses are likely to again use the single-stage pipeline for simplicity. Additional speed gains can be realized by moving the entire FINDCHIRP computation onto the GPU.

---

[2]If $\chi^2_{m,n}[j]$ is only required for one particular $j$ then there is a more efficient way to compute it. However, the FINDCHIRP algorithm does not employ the chi-squared test so much as a veto as a part of a constrained maximization of signal-to-noise for times in which the chi-squared condition is satisfied. Thus, $\chi^2_{m,n}[j]$ needs to be computed for all $j$ if it is computed at all.

A true signal in the data is expected to produce a sharp peak in the matched-filter output at almost exactly the correct termination time $t_0$ (usually within one sample point of the correct time in simulations). For sufficiently loud signals, however, a signal-to-noise threshold may be crossed for several samples even though the correct termination time will have a much greater signal-to-noise ratio than nearby times. Non-Gaussian noise artifacts may produce many threshold-crossing triggers, often for a duration similar to the duration of the inspiral template that is used. In principle, a large impulse in the detector output at sample $j_0$ can cause triggers for samples $j_0 - T_{\text{spec}}/\Delta t \leq j \leq j_0 + (T_{\text{spec}} + T_{\text{chirp},m})/\Delta t$. Rather than record triggers for all samples in which the signal-to-noise threshold is exceeded while the chi-squared test is satisfied, FINDCHIRP has the option of *maximizing over a chirp*: essentially clustering together triggers that lie within a time $T_{\text{chirp},m}$. Algorithmically, whenever $|\zeta_{m,n}[j]|^2 > \zeta_\star^2$ and $\Xi_{m,n}[j] < \Xi_\star$, a trigger is created with a value of $\rho$ and $\chi^2$. If this trigger is within a time $T_{\text{chirp},m}$ after an earlier trigger with a *larger* value of the signal-to-noise ratio $\rho$, discard the *current* trigger (it is clustered with the previous trigger). If this trigger is within a time $T_{\text{chirp},m}$ after an earlier trigger with a *smaller* signal-to-noise ratio $\rho$, discard the *earlier* trigger (the previous trigger is clustered with the current trigger). The result is a set of remaining triggers that are separated by a time of at least $T_{\text{chirp},m}$. Note that this algorithm depends on the order in which the triggers are selected, i.e., a different set of triggers may arise if the triggers are examined in inverse order of $j$ rather than in order of $j$. FINDCHIRP applies the conditions as $j$ is advanced from $j = N/4$ to $j = 3N/4 - 1$ (i.e., forward in time).[3] The effect of the maximizing over a chirp is to retain any true signal without introducing any significant bias in parameters, e.g., time of arrival (which can be demonstrated by simulations), while reducing the number of triggers that are produced by noise artifacts.

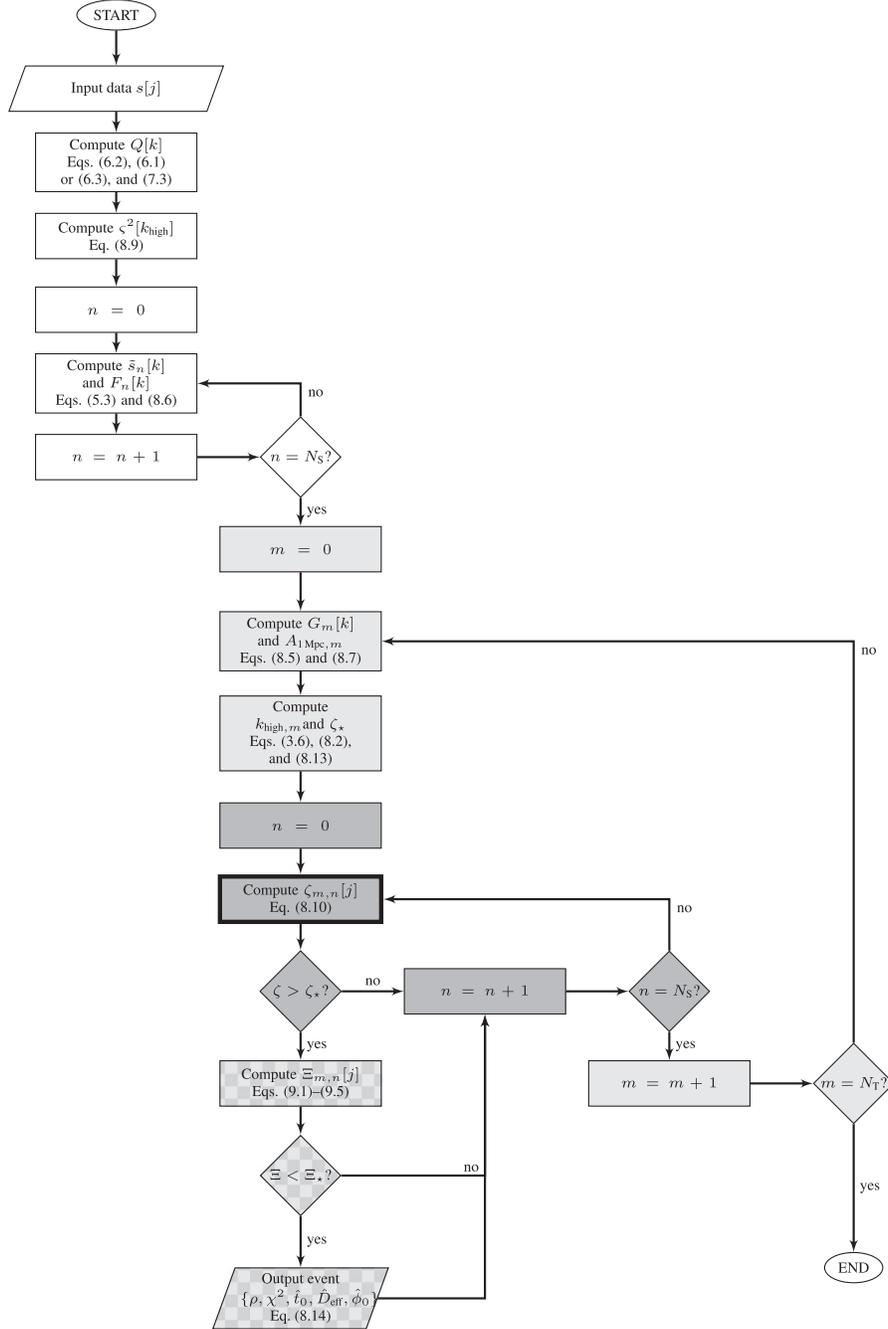

FIG. 1. FINDCHIRP flow chart. The main computation takes place in two loops: the outer loop (in light grey) is over templates while the inner loop (in dark grey) is over data segments. The chi-squared branch, shown with a checkerboard pattern, is within the inner loop and is computationally costly. However, this branch is only executed if a candidate event is identified. Therefore, the computational cost is dominated by the procedure shown with a thick border: the calculation of the matched filter given by Eq. (8.10).

---
[3]Other methods can also be employed, for example, maximizing all triggers that are separated in time by less than $T_{\text{chirp},m}$.

## XI. EXECUTION OF THE FINDCHIRP ALGORITHM

In this section, we describe the sequence of operations that comprise the FINDCHIRP algorithm and highlight the tunable parameters of each operation. The algorithm is also illustrated in Fig. 1.

Since we are only describing the FINDCHIRP algorithm itself, we assume that a bank of templates $(M, \mu)_m$ has already been constructed for a given minimal match $\delta$, according to the methods described in [8–10] and we are provided with either calibrated strain data $s[j]$ or the output error signal of the interferometer $e[j]$ and instrument calibration $R[k]$ as described in [48].

The initial operation is to scale the strain data by the dynamic-range factor $\kappa$, and divide it into data segments $\kappa s_n[j]$ suitable for analysis. The data segment duration $T$, stride length $\Delta$, and number of data segments $N_S$ in a block are selected. These quantities then define a data block length according to Eq. (5.2). A sample rate $1/\Delta t$ must be chosen (which must be less than or equal to the sample rate of the detector data acquisition system); the sample rate and data segment length define the number of points in a data segment $N = T/\Delta t$. As mentioned previously, the lengths and sample rate are chosen so that $N$ is an integer power of 2.

The next operation is construction of an average power spectrum $\kappa^2 S[k]$ using a specified data window $w[j]$ and power spectrum estimation method (Welch's method, the median method, or the median-mean method). The number of periodograms used in the average power spectrum estimate is typically chosen to be equal to the number of data segments, although different numbers of periodograms could be chosen. An inverse spectrum duration $T_{\text{spec}}$ is then given in order to construct the truncated inverse power spectrum $\kappa^{-2} Q[k]$, according to Eq. (7.3).

Each input data segment is Fourier transformed to obtain $\kappa \tilde{s}_n[k]$. The quantity $F_n[k]$ described in Eq. (8.6) can then be constructed. All frequency components of $F_n[k]$ below a specified low-frequency cutoff $f_{\text{low}}$ are set to zero, as are the DC and Nyquist components. The template-independent normalization constants $\varsigma[k_{\text{high}}]$ described in Eq. (8.9) are also computed at this point.

The algorithm now commences a loop over the $N_T$ templates in the bank, using the specified signal-to-noise-ratio and modified chi-squared thresholds, $\rho_\star$ and $\Xi_\star$, and the method of maximizing over triggers. For each template $(M, \mu)_m$, the FINDCHIRP template $G_m[k]$ is computed, according to Eq. (8.7). The high-frequency cutoff $k_{\text{high},m}$ for the template is obtained using Eq. (3.6) and used to select the correct value of $\varsigma^2[k_{\text{high},m}]$ for the template. The normalized signal-to-noise threshold $\zeta_\star$ is then computed for this template according to Eq. (8.13).

An inner loop over the FINDCHIRP data segments is then entered. For each FINDCHIRP segment $F_n[k]$ and FINDCHIRP template $G_m[k]$ the filter output $\zeta_{m,n}$ is computed according to Eq. (8.10). The trigger selection algorithm described in Sec. X is now used to determine if any triggers should be generated for this data segment and template, given the supplied thresholds and trigger maximization method. If necessary, the chi-squared veto is computed at this stage, according to Eq. (9.4) and the threshold given in Eq. (9.5). If any triggers are generated, the template parameters $(M, \mu)_m$ are stored, along with the termination time $\hat{t}_0$, signal-to-noise ratio, effective distance $\hat{D}_{\text{eff}}$, termination phase $\hat{\phi}_0$, chi-squared veto parameters, and the normalization constant $\sigma_m^2$ of the trigger. The triggers are generated and stored to disk for later stages of the analysis pipeline.

The segment index $n$ is then incremented and the loop over the data segments continues. Once all $N_S$ data segments have been filtered against the template, the template index $m$ is incremented and the loop over templates continues until all $N_T$ templates have been filtered against all $N_S$ data segments.

## XII. CONCLUSION

Profiling of the inspiral search code based on the FINDCHIRP algorithm was performed on a 3 GHz Pentium 4 CPU with seven data segments of length 256 seconds. The data was read from disk, resampled from 16384 Hz to 4096 Hz and filtered against a bank containing 474 templates using the FFTW package [50] to perform the discrete Fourier transforms; the resulting 1255 triggers were written out to disk. Of the 2909 seconds of execution time, 1088 seconds were spent performing complex FFTs required by the matched filter, and 1600 seconds performing the chi-squared veto. Of the time taken to perform the chi-squared veto, 1244 seconds are spent executing inverse FFTs. In total, 2300 seconds of the 2900 are spent doing FFTs, which means that the execution of the FINDCHIRP algorithm is FFT-dominated, as desired.

In practice, the FINDCHIRP algorithm is only a part of the search for gravitational waves from binary inspiral. An inspiral analysis pipeline typically includes data selection, template bank generation, trigger generation using FINDCHIRP, trigger coincidence tests between multiple detectors, vetoes based on instrumental behavior, coherent combination of the optimal filter output from multiple detectors, and finally manual candidate follow-ups. Pipelines vary between specific analyses and the topology of the analysis pipelines has evolved due to the specific demands of particular observing runs [14–16,18,19,21–25].

It is simple to modify the FINDCHIRP algorithm to use restricted post-Newtonian templates higher than second-order by adding addition terms to the construction of the FINDCHIRP template phase in Eq. (8.5b). The phasing equations used in the LSC implementation of FINDCHIRP are given in Appendix A for completeness. Appendix A described the modification of FINDCHIRP used to search for time-domain templates, e.g., those based on post-Newtonian resummation techniques, such as the effective one body formalism [37]. These modifications cannot make use of the factorization used in the stationary phase templates, but they allow efficient reuse of the search code developed and tested for the frequency-domain post-Newtonian templates.

## ACKNOWLEDGMENTS

We would like to thank Stanislav Babak for several useful suggestions and corrections on this work. This work has been supported in part by the National Science Foundation Grant Nos. PHY-9728704, PHY-0071028, PHY-0099568, PHY-0200852, PHY-0701817, PHY-0847611, and PHY-0970074, and by the LIGO Laboratory coperative agreement PHY-0107417. Patrick Brady is also grateful to the Alfred P Sloan Foundation for support. Patrick Brady and Duncan Brown also thank and the Research Corporation Cottrell Scholars Program for support.

## APPENDIX A: ALGORITHM FOR TEMPLATES GENERATED IN THE TIME-DOMAIN

The optimization of the FINDCHIRP algorithm described above is dependent on the use of frequency-domain-restricted post-Newtonian waveforms as the template. It is a simple matter, however, to modify the algorithm (and hence the code used to implement the algorithm) so that an arbitrary waveform generated in the time-domain $h(t)$ may be used as the matched-filter template. This allows use of inspiral templates such as the effective one body (EOB) [37] waveforms. These waveforms have been tuned to numerical relativity simulations of coalescing black holes and have a higher overlap with higher-mass signals in the sensitive band of the LIGO detectors [38]. Other waveforms of interest are described in [56–60]. In this Appendix, we describe the modifications necessary to use time-domain templates in FINDCHIRP.

We assume that the desired template waveform is generated in the form

$$h_{1\ \text{Mpc},m}(t) = A_m(t-t_0)\cos[2\phi_0 + 2\phi_m(t-t_0)], \quad \text{(A1)}$$

where $t_0$ and $\phi_0$ are the termination time and phase, as described in Sec. III, and $A_m(t)$ and $\phi_m(t)$ are the particular amplitude and phase evolution for the $m$th template in the bank. The bank may include parameterization over binary component spins as well as masses. The template waveform is generated from the low-frequency cutoff $f_{\text{low}}$ and is normalized to the canonical distance of 1 Mpc. Recall the factorization of the matched-filter output, given by Eq. (8.3):

$$\tilde{z}_{n,m}[k] = (\Delta f)^{-1} A_{1\ \text{Mpc},m} F_n[k] G_m[k]. \quad \text{(A2)}$$

Since we are now only provided with the numerical value of the waveform as a function of time, we cannot perform the same factorization of the waveform as for stationary phase templates. Instead, to compute the FINDCHIRP data segment $F_n[k]$, we remove the template-dependent amplitude by making the replacement $k^{-7/6} \to 1$ in Eq. (8.6). Similarly, the form of $A_{1\ \text{Mpc},m}$ is now much simpler, with $A_{1\ \text{Mpc},m} = 1$.

To construct the FINDCHIRP template $G_m[k]$, we construct a segment of length $N$ sample points and populate it with the discrete samples of the template waveform $h_{1\ \text{Mpc},m}[j]$. The waveform is sampled at the sampling interval of the matched filter $\Delta t$. When we place the waveform in this segment, we must ensure that the termination of the waveform is placed at the sample point $j = 0$, i.e., $t_0 = 0$. For example, if the template is a post-Newtonian waveform generated in the time-domain [40], then it is typical to end the waveform generation at the time which the frequency evolution of the waveform ceases to be monotonically-increasing and not the time at which the gravitational-wave frequency goes to infinity. Thus the last nonzero sample point of the generated template may *not* correspond to the formal time of coalescence. If the algorithm generating the waveform provides the FINDCHIRP algorithm a formal coalescence time, the waveform is placed so that the coalescence time is at the last sample point in the segment. Any subsequent signal (such as merger and ringdown, in the case of EOB) is wrapped around to the start of the segment. If a formal coalescence time is not provided, the waveform is placed so that the last nonzero sample of the template is at the end of the segment. As discussed in Sec. III, this causes some ambiguity in the meaning of the recorded arrival time, but as long as this is consistent among all detectors in the network, this is not a significant effect. After placing the waveform in the segment, we construct the discrete forward Fourier transform of the waveform, as described by Eq. (2.3) and construct

$$G_m[k] = \begin{cases} \Delta f \kappa \tilde{h}^*_{1\ \text{Mpc},m}[k] & k_{\text{low}} \leq k < k_{\text{high},m} \\ 0 & \text{otherwise.} \end{cases} \quad \text{(A3)}$$

Finally, we construct the normalization constant

$$\mathsf{s}_m^2 = 4\Delta f \sum_{k=k_{\text{low}}}^{k_{\text{high},m}-1} \kappa^{-2} Q[k] |\kappa \tilde{h}_{1\ \text{Mpc},m}[k]|^2, \quad \text{(A4)}$$

which is now dependent on the template parameters. Once we have constructed these quantities we may proceed with the FINDCHIRP algorithm described in Sec. VIII and IX to obtain the signal-to-noise ratio and the value of the chi-squared veto for the particular template we have chosen. The computational operations required per template are increased by $O(N \log N)$ for the additional real-to-half-complex forward FFT to construct $\tilde{h}_{1\ \text{Mpc},m}$, and $O(N)$ operations to construct $\mathsf{s}_m^2$.

## APPENDIX B: BIAS IN MEDIAN POWER SPECTRUM ESTIMATION

Here we compute the bias $\alpha$ of the median of a set of periodograms relative to the mean of a set of periodograms. We assume that the periodograms are obtained from Gaussian noise. In this Appendix, let us focus on one frequency bin of the periodogram, and for brevity we adopt the symbol $x$ for the power in the frequency bin, that is, we define $x_\ell = P_\ell[k]$ for $\ell = 1, \ldots, n$. (Here $n$ is the number of periodograms being averaged. It is either $N_S$ or $N_S/2$ depending on the choice of method.) Let $f(x)$ be the probability distribution function for $x$. For Gaussian noise, $f(x) = \mu^{-1} e^{-x/\mu}$ where

$$\mu = \langle x \rangle = \int_0^\infty x f(x) dx \tag{B1}$$

is the *population* mean of $x$. So $\mu = \langle P[k] \rangle$. The *population* median $x_{1/2}$ is defined by

$$\frac{1}{2} = \int_0^{x_{1/2}} f(x) dx \tag{B2}$$

which yields

$$x_{1/2} = \mu \ln 2. \tag{B3}$$

Thus the bias of the population median is $\alpha = \ln 2$.

The *sample* mean is unbiased compared to the population mean. The sample mean is

$$\bar{x} = \frac{1}{n} \sum_{\ell=1}^n x_\ell. \tag{B4}$$

The expected value of $\bar{x}$ is

$$\langle \bar{x} \rangle = \frac{1}{n} \sum_{\ell=1}^n \langle x_\ell \rangle = \mu \tag{B5}$$

so $\bar{x}$ is an unbiased estimator of $\mu$.

The *sample* median, however, does have a bias. The sample median is:

$$x_{\text{med}} = \text{median}\{x_\ell\}. \tag{B6}$$

To compute the bias, we first need to obtain the probability distribution for the sample median.

For simplicity, assume now that $n$ is odd. The probability of the sample median having a value between $x_{\text{med}}$ and $x_{\text{med}} + dx_{\text{med}}$ is proportional to the probability of one of the samples having a value between $x_{\text{med}}$ and $x_{\text{med}} + dx_{\text{med}}$ times the probability that half of the remaining samples are larger than $x_{\text{med}}$ and the other half are smaller than $x_{\text{med}}$. Thus, the probability distribution for $x_{\text{med}}$ is given by $g(x_{\text{med}})$ where

$$g(x_{\text{med}}) dx_{\text{med}} = C[1 - Q(x_{\text{med}})]^m Q^m(x_{\text{med}}) f(x_{\text{med}}) dx_{\text{med}} \tag{B7}$$

where $m = (n-1)/2$ is half the number of remaining samples after one has been selected as the median. Here, $Q(x)$ is the upper-tail probability of $x$, i.e., the probability that a sample exceeds the value $x$

$$Q(x) = \int_x^\infty f(x) dx = e^{-x/\mu}, \tag{B8}$$

where the second equality holds for the exponential distribution function corresponding to the power in Gaussian noise. The normalization factor $C$ is a combinatoric factor which arises from the number of ways of selecting a particular sample as the median and then choosing half of the remaining points to be greater than this value. Thus it has the value of $n$ (number of ways to select the median sample) times $n - 1 = 2m$ choose $(n-1)/2 = m$ (number of ways of choosing half the points to be larger)

$$C = n \times \binom{2m}{m} = \frac{1}{B(m+1, m+1)}, \tag{B9}$$

where $B(x, y) = \int_0^1 t^{x-1}(1-t)^{y-1} dt$ is the Euler beta function. This factor can also be obtained simply by normalizing the probability distribution $g(x_{\text{med}})$. To do so it is useful to make the substitution $t = Q(x_{\text{med}})$ so that $dt = -f(x_{\text{med}}) dx_{\text{med}}$ and

$$1 = \int g(x) dx = \int_0^1 C t^m (1-t)^m dt$$
$$= CB(m+1, m+1). \tag{B10}$$

Now we can compute the expected value for the median. Note that for the exponential probability distribution, $\langle x_{\text{med}} \rangle = -\mu \ln t$. Thus $x_{\text{med}} = -C\mu \int_0^1 t^m (1-t)^m \ln t \, dt$. To evaluate the integral, start with the definition (above) of the Euler beta function. Differentiate it with regard to the first argument, and use the definition of the digamma function $\psi(z) = \Gamma'(z)/\Gamma(z)$ to obtain

$$\langle x_{\text{med}} \rangle = [\psi(2m+2) - \psi(m+1)]\mu = \mu \sum_{\ell=1}^n \frac{(-1)^{\ell+1}}{\ell} \tag{B11}$$

where we have used Eqs. 8.365.6, 8.365.2, and 8.375.2 of Ref. [66] to reduce $\psi$. The bias factor is therefore

$$\alpha = \sum_{\ell=1}^n \frac{(-1)^{\ell+1}}{\ell} \tag{B12}$$

for odd $n$. This result makes sense: As $n \to \infty$ the series approaches $\ln 2$ which is the bias for the population median. However, for $n = 1$, $\alpha = 1$, so there is no bias (the median is equal to the mean for one sample!).

## APPENDIX C: CHI-SQUARED STATISTIC FOR A MISMATCHED SIGNAL

For simplicity we write the chi-squared statistic [13] in the equivalent form [cf. Eq. (7.1)]

$$\chi^2[j] = \sum_{\ell=1}^p \frac{|z_\ell[j] - z[j]/p|^2}{\sigma^2/p}, \tag{C1}$$

where

$$z_\ell[j] = 4\Delta f \sum_{k=k_{\ell-1}}^{k_\ell - 1} \frac{\tilde{s}[k] \tilde{h}^*_{1 \text{ Mpc}}[k]}{S[k]} e^{2\pi i j k/N}, \tag{C2}$$

$$z[j] = \sum_{\ell=1}^{p} z_\ell[j], \quad\quad (C3)$$

and

$$\sigma^2 = 4\Delta f \sum_{k=1}^{\lfloor (N-1)/2 \rfloor} \frac{|\tilde{h}_{1\ \text{Mpc}}[k]|^2}{S[k]}. \quad\quad (C4)$$

For brevity we have dropped the indices $n$ and $m$; the explicit dependence on $j$ will also be dropped hereafter. In this Appendix we further simplify the notation by adopting normalized templates $\tilde{u}[k] = \tilde{h}_{1\ \text{Mpc}}[k]/\sigma$. In terms of these templates we define the inner products as in [13]

$$(s, u)_\ell = 4\Delta f \sum_{k=k_{(\ell-1)}}^{(k_\ell - 1)} \frac{\tilde{s}[k]\tilde{u}^*[k]}{S[k]} e^{2\pi i j k/N} \quad\quad (C5)$$

for the $p$ different bands, which are chosen so that $(u, u)_\ell = 1/p$, and the inner product

$$(s, u) = \sum_{\ell=1}^{p} (s, u)_\ell = 4\Delta f \sum_{k=1}^{\lfloor (N-1)/2 \rfloor} \frac{\tilde{s}[k]\tilde{u}^*[k]}{S[k]}. \quad\quad (C6)$$

With this notation, the signal-to-noise ratio is given by $\rho^2 = |(s, u)|^2$ and chi-squared statistic is

$$\chi^2 = \sum_{\ell=1}^{p} \frac{|(s, u)_\ell - (s, u)/p|^2}{1/p}$$

$$= -|(s, u)|^2 + p \sum_{\ell=1}^{p} |(s, u)_\ell|^2. \quad\quad (C7)$$

To see how the chi-squared statistic is affected by a strong signal (considerably larger than the noise)[4], suppose that the detector output $s[j]$ consists of the gravitational waveform $av[j]$ where $a$ specifies the amplitude of the gravitational-wave. Here $v[j]$ is also a normalized [in terms of the inner product of Eq. (C6)] gravitational waveform that is not exactly the same as $u[j]$. The discrepancy between the two waveforms is given by the *mismatch*

$$\delta = 1 - |(v, u)|. \quad\quad (C8)$$

The mismatch is the fraction of the signal-to-noise ratio that is lost by filtering the true signal $av[j]$ with the template $u[j]$ compared to if the template $v[j]$ were used. The chi-squared statistic is

$$\chi^2 = -a^2|(v, u)|^2 + pa^2 \sum_{\ell=1}^{p} |(v, u)_\ell|^2$$

$$\leq -a^2|(v, u)|^2 + pa^2 \sum_{\ell=1}^{p} (v, v)_\ell (u, u)_\ell$$

$$= -\rho^2 + a^2 \leq 2a^2\delta \approx 2\rho^2\delta \quad\quad (C9)$$

---

[4]The strong-signal assumption holds if $p - 1 \ll 2a^2\delta$. This can be seen from a full calculation, cf. Eq. (6.24) of Ref. [13].

where we have used the Schwarz inequality to obtain the second line and the normalization condition $(u, u)_\ell = 1/p$ to obtain the third. The final approximation assumes that $\delta \ll 1$. Thus, the chi-squared statistic is offset by an amount that is bounded by twice the squared signal-to-noise ratio observed times the mismatch factor. There is no offset for a template that perfectly matched the signal waveform.

It can be shown [13] that in the presence of a signal and Gaussian noise that $\chi^2$ has a noncentral chi-squared distribution [62] with $\nu = 2p - 2$ degrees of freedom and a noncentral parameter $\lambda \lesssim 2\rho^2\delta$ (where now $\lambda$ might possibly be slightly greater than $2\delta$ times the *measured* signal-to-noise ratio squared owing to the presence of the noise). This distribution has a mean value of $\nu + \lambda$ and a variance of $2\nu + 4\lambda$. We see then that the modified chi-squared statistic $\Xi$ of Eq. (9.5) has a mean of $\lesssim 2$ and a variance of $\lesssim (4 \text{ or } 8)/(p + \rho^2\delta)$ (4 when $p \gg \rho^2\delta$ and 8 when $p \ll \rho^2\delta$) for Gaussian noise. Thus we would expect to set a threshold on $\Xi$ of $\Xi_\star \sim$ a few.

## APPENDIX D: DISTANCE CONVENTIONS

The *effective distance*, $D_{\text{eff}}$, is a measure of the inverse amplitude of a gravitational-wave signal produced in a detector. For a binary that is optimally located (directly above or below a two-arm interferometric detector) and oriented (where the binary's orbit is viewed face-on), the effective distance is equal to the *actual distance*, $D$. For a source that is not optimally located or oriented, the signal is weaker than one from an optimally located and oriented source at the same actual distance, and therefore the effective distance is greater than the actual distance.

If a signal from a system having a total mass $M$, a reduced mass $\mu$, and an effective distance $D_{\text{eff}}$ is present in a detector's data stream, so that it induces a strain $h(t)$ on the detector, then the *expected* signal-to-noise ratio found for a matched-filter search is

$$\varrho = \sqrt{4 \int_0^\infty \frac{|\tilde{h}(f)|^2}{S_n(f)} df}$$

$$= \left(\frac{1\ \text{Mpc}}{D_{\text{eff}}}\right) \sqrt{4 \mathcal{A}_{1\ \text{Mpc}}^2 (M, \mu) \int_0^\infty \frac{f^{-7/3}}{S_n(f)} df} \quad\quad (D1)$$

if the data is filtered with a template that has exactly the same waveform parameters as the signal. A common source of confusion is the use of the term "signal-to-noise ratio" in reference to the matched-filter output for a particular template, $\rho_m(t)$, as well as in reference to the measure of the strength of a signal relative to a detector's noise, $\varrho$. The latter should be called the *expected signal-to-noise ratio*, while the actual observed signal-to-noise ratio will differ owing to the presence of random noise as well as the possible mismatch between the true signal and the template used in filtering.

The effective distance is related to other distances that have become widely used in gravitational-wave physics. The *horizon distance*, $D_{\mathrm{hor}}$, is the effective distance of a particular source, a $1.4 M_\odot$–$1.4 M_\odot$ binary neutron star system, that would produce an expected signal-to-noise ratio $\varrho = 8$. That is, the horizon distance is defined by

$$D_{\mathrm{hor}} = \frac{1 \text{ Mpc}}{\varrho} \sqrt{4 \mathcal{A}_{1 \text{ Mpc}}^2 (M, \mu) \int_0^\infty \frac{f^{-7/3}}{S_n(f)} df}, \quad \text{(D2)}$$

where $\varrho = 8$, $M = 2.8 M_\odot$, and $\mu = 0.7 M_\odot$. Horizon distance is therefore a measure of the sensitivity of a detector: it depends on the integral of the inverse power spectral density with a frequency weighting of $f^{-7/3}$.

The *sense-monitor range* [63], $\mathcal{R}$, is related to the horizon distance by $\mathcal{R} = \mathcal{F} D_{\mathrm{hor}}$ where $\mathcal{F}^{-1} \simeq 2.2648$.[5] It is called sense-monitor range after the data monitor SENSEMONITOR [64] that is run in the LIGO control rooms. The sense-monitor range represents not the farthest distance to which a detector is sensitive, but, rather, the radius of a sphere that would contain as many sources as the number that would produce an expected signal-to-noise ratio $\varrho = 8$ in a detector under the assumption of a homogeneous distribution of sources having random orientations. Specifically, suppose that there are $N_{\mathrm{hor}}$ sources homogeneously distributed in a sphere of radius $D_{\mathrm{hor}}$. Each of these sources has an effective distance $D_{\mathrm{eff}}$ (which is always greater than or equal to its actual distance $D$). The number $N_{\mathrm{det}}$ of these sources that are detectable (i.e., which would produce an expected signal-to-noise ratio $\varrho = 8$) is simply the number of the sources that have $D_{\mathrm{eff}} \leq D_{\mathrm{hor}}$; the quantity $\mathcal{F}^3 = N_{\mathrm{det}}/N_{\mathrm{hor}}$ is the fraction of the universe within the horizon distance that is actually probed by the detector. (Note that $\mathcal{F}$ is *not* equal to $5^{-1/2}$, as it is often mistaken to be.)

While the actual distance, $D$, is a physical parameter of a source, the effective distance $D_{\mathrm{eff}}$ is related to the inverse of the strain amplitude seen in a detector. The horizon distance $D_{\mathrm{hor}}$ and sense-monitor range $\mathcal{R}$, on the other hand, are not related to an actual source but rather are measures of the sensitivity of the detector. The horizon distance is a representation of the maximum distance to which a detector can see, while the sense-monitor range describes the volume that may be surveyed by a detector.

Finally, Ref. [65] defines a distance quantity which is useful in the computation of rate limits on coalescing binaries [43]. If an inspiral signal observed in the detector does not terminate in the sensitive band of the detector (as is the case for $M \lesssim 10 M_\odot$ binaries in Initial LIGO), then it follows from Eq. (3.4b) that the amplitude of the waveform is proportional to $\mathcal{M}^{-5/6} D_{\mathrm{eff}}$. For the purpose of computing the efficiency of a search, it is convenient to define the *chirp distance*

$$D_{\mathrm{c}} = \left(\frac{\mathcal{M}_{\mathrm{BNS}}}{\mathcal{M}}\right)^{-5/6} D_{\mathrm{eff}}, \quad \text{(D3)}$$

where $\mathcal{M}_{\mathrm{BNS}}$ is the chirp mass of a binary with $m_1 = m_2 = 1.4 M_\odot$.

## APPENDIX E: EXTENSION TO HIGHER POST-NEWTONIAN ORDERS

Extending to the FINDCHIRP algorithm to use higher post-Newtonian orders is straighforward. At 3.5 post-Newtonian order, the expression for the chirp time given by Eq. (3.5a) becomes (see Eq. (3.8b) of Ref. [36])

$$\begin{aligned}
T_{\mathrm{chirp}} = \frac{5}{256\eta} \frac{GM}{c^3} \Bigg\{ & v_{\mathrm{low}}^{-8} + \left(\frac{743}{252} + \frac{11}{3}\eta\right) v_{\mathrm{low}}^{-6} - \frac{32}{5}\pi v_{\mathrm{low}}^{-5} + \left(\frac{3058673}{508032} + \frac{5429}{504}\eta + \frac{617}{72}\eta^2\right) v_{\mathrm{low}}^{-4} + \left(\frac{13}{3}\eta - \frac{7729}{252}\right) \pi v_{\mathrm{low}}^{-3} \\
& + \bigg[\frac{6848 \gamma_E}{105} - \frac{10052469856691}{23471078400} + \frac{128\pi^2}{3} + \left(\frac{3147553127}{3048192} - \frac{451\pi^2}{12}\right)\eta - \frac{15211}{1728}\eta^2 + \frac{25565}{1296}\eta^3 + \frac{6848}{105} \ln(4 v_{\mathrm{low}})\bigg] v_{\mathrm{low}}^{-2} \\
& + \left(\frac{14809}{378}\eta^2 - \frac{75703}{756}\eta - \frac{15419335}{127008}\right) \pi v_{\mathrm{low}}^{-1}\Bigg\}
\end{aligned} \quad \text{(E1)}$$

and the stationary phase approximation to 3.5 post-Newtonian order is given by Eqs. (3.4a) and (3.4b) with the stationary phase function extended to include higher post-Newtonian terms (see Eq. (3.18) of Ref. [36])

$$\begin{aligned}
\Psi(f; M, \mu) = 2\pi f t_0 - 2\phi_0 - \frac{\pi}{4} + \frac{3}{128}\frac{1}{\eta} v^{-5} \Bigg\{ & 1 + \left(\frac{3715}{756} + \frac{55}{9}\eta\right) v^2 - 16\pi v^3 + \left(\frac{15293365}{508032} + \frac{27145}{504}\eta + \frac{3085}{72}\eta^2\right) v^4 \\
& + \left(\frac{38645}{756} - \frac{65}{9}\eta\right)\left[1 + 3\ln\left(\frac{v}{v_0}\right)\right] \pi v^5 + \bigg[\frac{11583231236531}{4694215680} - \frac{640}{3}\pi^2 - \frac{6848}{21}\gamma_E - \frac{6848}{21}\ln(4v) \\
& + \left(-\frac{15737765635}{3048192} + \frac{2255}{12}\pi^2\right)\eta + \frac{76055}{1728}\eta^2 - \frac{127825}{1296}\eta^3\bigg] v^6 + \left(\frac{77096675}{254016} + \frac{378515}{1512}\eta - \frac{74045}{756}\eta^2\right) \pi v^7 \Bigg\},
\end{aligned} \quad \text{(E2)}$$

---

[5]The factor 2.2648 is equal to $4/(3 \times 1.84)^{1/3}$ where 1.84 is the value in Eq. (5.3) of [63].

where $\gamma_E \approx 0.577216$ is the Euler constant, and $\nu$ is the post-Newtonian parameter given by Eq. (3.4d). Here $\nu_0$ is an arbitrary reference value for the post-Newtonian parameter, which is normally taken to be $\nu_0 = 1$.